\documentclass[12pt,preprint]{aastex}

\newcommand{\bq}{\begin{equation}}
\newcommand{\eq}{\end{equation}}

\shorttitle{A Massive, Post-Starburst Galaxy at $z\sim 6.6$?}
\shortauthors{Mobasher {\it et al}}

\begin{document}

%\title{A Likely Massive Post-Starburst Galaxy at $\mathbf{z \sim 6.6}$}
%\newline [*** POSSIBLE NEW TITLE ***]

\title{Evidence for a Massive Post-Starburst Galaxy at $\mathbf{z \sim 6.5}$}

\author{B. Mobasher\altaffilmark{1,2}, M. Dickinson\altaffilmark{3},
H. C. Ferguson\altaffilmark{1}, M. Giavalisco\altaffilmark{1},
T. Wiklind\altaffilmark{1,2}, 
D. Stark\altaffilmark{4}, 
R. S. Ellis\altaffilmark{4}, M. Fall\altaffilmark{1},
N. A. Grogin\altaffilmark{5}, L. A. Moustakas\altaffilmark{1},
N. Panagia\altaffilmark{1,2}, 
M. Sosey\altaffilmark{1}, 
M. Stiavelli\altaffilmark{1}, E. Bergeron\altaffilmark{1}, 
S. Casertano\altaffilmark{1}, P. Ingraham\altaffilmark{6}
A. Koekemoer\altaffilmark{1}, I. Labb\'e\altaffilmark{7}
M. Livio\altaffilmark{1},
B. Rodgers\altaffilmark{6}, C. Scarlata\altaffilmark{8}, 
J. Vernet\altaffilmark{9}, A. Renzini\altaffilmark{9,10}, 
P. Rosati\altaffilmark{9}, H. K\"{u}ntschner\altaffilmark{11}, 
M. K\"{u}mmel\altaffilmark{11} ,  J. R. Walsh\altaffilmark{11}, 
R. Chary\altaffilmark{12},
P. Eisenhardt\altaffilmark{13}, D. Stern\altaffilmark{13}}

%\newline {\bf [*** UPDATE AUTHORS ***]} } 
\altaffiltext{1}{Space Telescope Science
Institute, 3700 San Martin Drive, Baltimore MD 21218, USA;
b.mobasher@stsci.edu} \altaffiltext{2}{Affiliated with the Space Sciences
Department of the European Space Agency} \altaffiltext{3}{NOAO, 950 N. Cherry
Ave. P.O. 26732, Tucson, AZ 85726-6732, USA} \altaffiltext{4}{105-24
Astronomy, Caltech, Pasadena, CA 91125 USA} \altaffiltext{5}{Johns Hopkins
University, 3400 N. Charles St. Baltimore, MD 21218, USA} 
\altaffiltext{6}{Gemini Observatory, Casilla 603, La Serena, Chile}
\altaffiltext{7}{The Carnegie Observatories, 813 Santa Barbara Street, Pasadena, CA 91101-1292}
\altaffiltext{8}{Institute fur Astronomie, ETH Zurich, HPF D8 Honggerberg, CH-8099 Zurich, Switzerland}
\altaffiltext{9}{European Southern Observatory, Karl Schwarzschild Strasse 2,
D-85748, Garching bei Munchen, Germany}
\altaffiltext{10}{INAF, Osservatorio Astronomico di Padova, Vicolo 
dell'Osservatorio 5, I-35122 Padova, Italy}
\altaffiltext{11}{Space Telescope European Coordinating Facility, European Southern Observatory, Karl Schwarzschild Strasse 2,
D-85748, Garching bei Munchen, Germany}
\altaffiltext{12}{Spitzer Science Center, California Institute of Technology, Mail Stop 220-6, Pasadena, CA 91125}
\altaffiltext{13}{Jet Propulsion Laboratory, California Institute of Technology, Mail Stop 169-506, Pasadena, CA91109}

\begin{abstract}
   We describe results from a search for high-redshift $J$--band
   ``dropout'' galaxies in the portion of the Great Observatories
   Origins Deep Survey (GOODS) southern field that is covered
   by extremely deep imaging from the Hubble Ultradeep Field (HUDF).
   Using observations at optical, near-infrared and mid-infrared
   wavelengths from the Hubble and Spitzer Space Telescopes and the
   European Southern Observatory Very Large Telescope, we find one
   particularly remarkable candidate, which we designate as HUDF-JD2.
   Its spectral energy distribution has distinctive features that are
   consistent with those of a galaxy at $z \sim 6.5$, observed several
   hundred million years after a powerful burst of star formation
   that produced a stellar mass of $6\times 10^{11} M_\odot$
   (for a Salpeter IMF).  We interpret a prominent photometric break
   between the near-infrared and Spitzer bandpasses as the 3646\AA\
   Balmer discontinuity.  The best-fitting models have low reddening
   and ages of several hundred Myr, placing the formation of the
   bulk of the stars at $z > 9$.  Alternative models of dusty
   galaxies at $z \approx 2.5$ are possible but provide significantly
   poorer fits to the photometric data.  The object is detected
   with Spitzer at 24$\mu$m.  We consider interpretations of the
   24$\mu$m emission as originating from either an obscured active
   nucleus or from star formation, and find that the 24$\mu$m
   detection does not help to uniquely discreminate between the
   $z = 6.5$ and 2.5 alternatives.   We present optical and
   near-infrared spectroscopy which has, thus far, failed to detect
   any spectral features.  This non-detection helps limit the
   solution in which the galaxy is a starburst or active galaxy at
   $z \approx 2.5$, but does not rule it out.  If the high-redshift
   interpretation is correct, HUDF-JD2 is an example of a galaxy that
   formed by a process strongly resembling traditional models of
   monolithic collapse, in a way which a very large mass of stars formed
   within a remarkably short period of time, at very high redshift.

\end{abstract}


\section{Introduction}

An important goal of observational cosmology is to measure the early
history of star formation and stellar mass assembly in galaxies.
Such investigations are key elements for testing galaxy formation 
theories (e.g. Somerville et al.\ 2001) and the early history of
star formation at $z > 6$, providing critical constraints on the time 
evolution of cosmic re--ionization (Loeb \& Barkana 2002; 
Stiavelli, Fall \& Panagia 2004).  Ultraviolet--bright, star-forming 
Lyman break galaxies (LBGs) have been studied in great 
detail at $z\sim 3$ and 4 by means of spectroscopic surveys (Steidel et al.\ 2003), 
and properties such as their star formation rates, stellar masses, ages,
and morphologies have been empirically characterized (see Giavalisco 2002 for
a review).  Galaxies at $z \approx 2$ to 3 with redder spectral energy 
distributions (SEDs) and large stellar masses 
($\sim 10^{11} M_\odot$) have also been identified via infrared 
surveys (e.g., Franx et al.\ 2003;  van Dokkum et al.\ 2003; 
Daddi et al.\ 2004; Yan et al.\ 2004).  A subset of these galaxies 
appear to be dominated by old stellar populations 
(e.g., Labb\'e et al.\ 2005), suggesting that some 
high redshift galaxies formed a large fraction of their stars at
substantially higher redshifts, perhaps $z > 5$.  Ever--increasing
numbers of star-forming galaxies at $4 < z \lesssim\ 6.5$ are being
identified in new (mostly optical) surveys, mainly via the Lyman 
break (Giavalisco et al.\ 2004a; Ouchi et al.\ 2004) and 
narrow-band (Rhoads et al.\ 2004; Ajiki et al.\ 2004) techniques.
The recently acquired Hubble Ultra Deep Field (HUDF) observations, 
which are the deepest images of the Universe at optical 
and near-infrared wavelengths, have been used to 
extend color--selected searches to $z > 7$ (Bunker et al.\ 2004; 
Yan \& Windhorst 2004; Bouwens et al.\ 2004), and have identified
a number of extremely faint candidates which await further studies.

The HUDF is situated within the southern field of the {\it Great 
Observatories Origins Deep Survey (GOODS)} (Dickinson \& Giavalisco 2003;  
Giavalisco et al.\ 2004b).  Here, we combine the HUDF HST images 
at 0.44 to 1.6$\mu$m with GOODS data in the $K_s$--band (2.16$\mu$m) 
and from the {\it Spitzer Space Telescope} at 3.6 to 24$\mu$m,
to search for candidate ``$J$-band dropouts'', i.e., objects whose 
light at 1$\mu$m (observed wavelength) is suppressed by the hydrogen 
opacity of the intergalactic medium (IGM) at $z \gtrsim 7$.  
We adopt $H_0=70$ km s$^{-1}$ Mpc$^{-1}$, $\Omega_{m}=0.3$ and 
$\Omega_{\Lambda}=0.7$ throughout this paper. All magnitudes are 
in the AB system.

\section{Imaging data}

The HUDF was imaged at optical wavelengths with the Advanced Camera 
for Surveys (ACS) Wide Field Camera (WFC), which covers a field of
$3\arcmin \times 3\arcmin$.  The images were taken in four passbands: 
F435W ($B_{435}$), F606W ($V_{606}$), F775W ($i_{775}$), and F850LP 
($z_{850}$).  These are the deepest optical images ever obtained, 
with limiting sensitivity of 29.3 to 28.5 AB magnitudes in the four 
filters ($10\sigma$ for extended source measured over 0.2 arcsec$^2$ 
aperture).  A detailed description of the HUDF ACS observations and 
data reduction will be presented in Beckwith et al.\ (in preparation).

Near-infrared observations of the HUDF observations were carried out with the 
Near-Infrared Camera and Multiobject Spectrograph (NICMOS), and cover a field 
of $2\farcm5\times 2\farcm5$ with the F110W ($J_{110}$) and F160W ($H_{160}$) 
filters (Thompson et al.\ 2005).  We have used an independent reduction of 
the HUDF NICMOS data made at STScI (Robberto et al., in preparation), 
which has slightly reduced noise and has fewer pixel artifacts compared to 
the Version 1 public release data 
products.\footnote{These have since been replaced by improved Version 2 
NICMOS data products.} The NICMOS images in both bands reach sensitivities 
of 26.9 AB magnitudes ($10\sigma$ in an aperture diameter 
of 0\farcs6). GOODS near-infrared data also include a mosaic of $K_s$--band 
images from the VLT/ISAAC instrument which covers the HUDF region 
with typical exposure times of 6 hours per pointing, seeing 
FWHM~$\approx 0\farcs4$, and limiting sensitivity $K_s = 25.1$ 
($10\sigma$ point source in a $1\arcsec$ diameter aperture; 
Vandame et al.\ in preparation; Giavalisco et al.\ 2004b).  

 The GOODS Legacy program (Dickinson et al., in preparation) with the 
 Spitzer Space Telescope has also surveyed the HUDF using the Infrared 
 Array Camera (IRAC) and the Multiband Imaging Photometer for Spitzer (MIPS).
 The IRAC data consist of images in four bands (3.6, 4.5, 5.8 and 8.0$\mu$m).
 The HUDF is located in an overlap strip which is covered twice by IRAC in 
 observing epochs separated by six months, and has therefore received 
 approximately 46 hours of exposure per channel.  The formal $10\sigma$ point 
 source limits range from 25.8 to 23.0 AB mag in the four channels, although 
 in practice, measurement errors at the faintest fluxes are mainly due to 
 crowding from neighboring objects.  The object discussed in this paper is 
 well above the faint sensitivity limit of the GOODS IRAC images.  
 The MIPS observations at 24$\mu$m have an integration time of approximately 
 10.4 hours, and reach a 5$\sigma$ limiting sensitivity of approximately 
 20$\mu$Jy.  Source extraction and catalogs for the GOODS 24$\mu$m data 
 will be described in Chary et al.\ (in preparation).

\section{Identification and photometry of candidate $J$-band dropout galaxies}

We generated a multi-wavelength source catalog based on detections made in
the NICMOS F160W band.  We used SExtractor (Bertin \& Arnout 1996) in 
dual-image mode with the parameters (kernel, number of
connected pixels, threshold) optimised by means of simulations. 
These consisted of placing artificial galaxies with known observable 
parameters in the images and retrieving them following the same procedures 
used for the real galaxies.  We used these catalogs to select very red 
galaxies 
with $(J_{110}-H_{160}) > 1.3$ and no detection (at the 2$\sigma$ level) 
at wavelengths shorter than $J_{110}$, down to the limit of the ACS/HUDF 
images (i.e. $J$-band dropouts). The adopted value of $J_{110}-H_{160}$ is
the observed color of an LBG with typical rest--frame UV colors, observed 
at $z=8$ and taking into account the appropriate cosmic opacity.

Two sources were found that satisfy these criteria:  UDF033242.88-274809.5 and \hfil\break
UDF033238.74-274839.9 (J2000 coordinates).  Henceforth, we will refer 
to these objects as HUDF-JD1 and HUDF-JD2 respectively. Both objects are well 
detected in the ISAAC $K_s$-band and in all four IRAC channels.  HUDF-JD2 
is clearly resolved in the NICMOS images relative to the point spread 
function (PSF); we estimate its half-light radius to be $\simeq0.30\pm$0.05
arcsec.  HUDF-JD1 is more compact, but is probably extended as well.  
 Although HUDF-JD1 statisfies the $(J_{110} - H_{160}) > 1.3$ color
 criterion and initially passed the optical non-detection test as well,
 deeper HUDF/ACS catalogs, generated later, detect faint $z$--band
 emission at this position.  No such optical emission is detected 
 for HUDF-JD2:  measurements within an $0.9$ arcsec diameter aperture 
 are consistent (within $\pm 2\sigma$) with zero flux in all four 
 ACS bands.  We also combined the $BViz$ ACS images to obtain a 
 ``white light'' image with a $3\sigma$ detection limit of 31.5 mag 
 over a $1.5$ arcsec diameter aperture.  HUDF-JD2 is still undetected 
 in this combined image, nor is it visible when the combined ACS image 
 is convolved to match the NICMOS PSF.

Both objects were noted previously by Chen \& Marzke (2004) and by 
Yan et al.\ (2004), using different selection criteria.  
Prior to the availability of the Spitzer data, which we will demonstrate
is crucial in the interpretation, Chen \& Marzke used the NICMOS and ACS data 
alone and identified these objects as candidates for evolved galaxies with 
photometric redshifts $z \approx 2.9$ and $4.25$, respectively.    
Yan et al.\ (2004) used the GOODS IRAC data to identify 17 
HUDF objects with extremely red ACS to IRAC colors (``IRAC Extremely Red 
Objects''). They estimated $z\approx 3.6$ and 3.4 for the two objects 
discussed here; however, they considered the photometric redshifts and 
stellar population modeling to be uncertain due to the lack of any optical 
detection, and therefore excluded these two objects from the detailed 
analysis of their sample.

The observed SED of HUDF-JD1 rises smoothly longward of the J-band,  
in contrast to the HUDF-JD2 SED which is almost flat in the $H/K$ region 
(i.e. a blue $H-K$ color),  followed by a significant break in between the 
$K_s$ and $3.6\ \mu m$ wavelengths (i.e. a red $K-3.6$ color). We will argue 
below that this distinctive feature, seen only in HUDF-J2,  is strong 
evidence for a high redshift galaxy ($z\sim 7$). By contrast, the smooth  SED 
of  HUDF-JD1 is consistent with  a dusty post-starburst at $z=2.4$ with 
E$_{\mathrm B-V} = 0.8$, an age of 200 Myr and metallicity $Z = 0.008$.  The  
two sources differ in other respects. X-ray observations of the GOODS-S field 
(the 1 Msec X-ray survey) with the Chandra X-ray Observatory (Giacconi et al.\ 2002) 
shows HUDF-JD1 to be coincident (within 1 arcsec) with a faint X-ray source 
(D.\ Alexander, private communication, based on an extension of the faint 
supplemental X-ray catalog from Alexander et al.\ 2003).  HUDF-JD2, 
instead, has no significant X-ray detection, with $3\sigma$ 
 upper limits of $5.1 \times 10^{-17}$ and $2.4\times 10^{-16}$ erg~s$^2$~cm$^{-2}$
in the soft (0.5-2~keV) and hard (2-8~keV) bands, respectively.    
For the remainder of this paper, 
we will concentrate on interpreting HUDF-JD2.  Images of HUDF-JD2 from 
0.44 to  8.0$\mu$m are shown in Figure 1.

We carefully remeasured photometry for HUDF-JD2 in each imaging data set 
through a variety of apertures, and using different sky background estimates, 
paying close attention to the possible contaminating effects of nearby 
neighbors.  There are three faint, blue objects, visible at wavelengths
as short as the ACS $B_{435}$--band, located slightly more than 1 arcsec 
from the $H_{160}$--band centroid of HUDF-JD2.   For large photometric  
apertures, these neighbors will significantly affect measurements in 
the NICMOS $J_{110}$--band or in ACS, but they have relatively small 
effects in the $H_{160}$ band and at longer wavelengths.  We have masked 
circular regions with diameter 0.5 arcsec around each of these sources 
before measuring photometry in the NICMOS and ISAAC images, and use 
a smaller aperture which excludes them in ACS.

A portion of the spectral energy distribution of HUDF-JD2 which will feature
prominently in our later discussion is the  ``break'' seen between the 
$K_s$ and 3.6$\mu$m IRAC bands.   Since the signal--to--noise ratio ($S/N$) 
in the GOODS ISAAC $K_s$-band image is less than that in the NICMOS 
$H_{160}$ or IRAC images, it is important to verify that this is
not an artifact due to measurement error.  Here we use a deeper (14 hour 
exposure time) VLT/ISAAC $K_s$ band image of the HUDF 
(Labb\'e et al.\, in preparation).  We calibrated the
photometric zeropoint of the deeper image to that of the GOODS
$K_s$--band data by comparing measurements for 8 galaxies
common to the two images.  Photometry was then repeated on
the deep $K_s$ image, and we found excellent agreement
with the measurement from the GOODS images within the 
nominal photometric uncertainties of the two data sets.
Because of its higher $S/N$, we use the
$K_s$ measurements from the deeper image here.

We measured the $JHK$ band magnitudes through circular 
apertures with diameters as large as 3 arcsec, and found 
that the photometric curve of growth converges by a diameter 
of 2 arcsec.  We examined the dependence of aperture 
magnitudes on different prescriptions for sky subtraction 
(concentric annuli or median of the data in a local region) 
and found that for aperture diameters of $2\arcsec$ and 
smaller, these did not affect the measurements at a level 
greater than our estimates for the photometric uncertainty 
due to background shot noise.  For the $JHK$ photometry, 
we choose 2\arcsec\ aperture diameters as a trade-off between 
convergence on the total flux, net $S/N$, and the effects 
of systematic uncertainty in the background subtraction.  
We also verified the large--aperture $J_{110}-H_{160}$ 
and $H_{160}-K_s$ colors by comparing them to measurements
through a smaller (0\farcs9 diameter) aperture using 
versions of the NICMOS images which were carefully convolved 
to match the PSF of the ISAAC $K_s$ image using a kernel 
derived by Fourier techniques.  The small--aperture measurements
have higher $S/N$ and are less subject to sky subtraction 
errors, and the results were in excellent agreement with the 
large--aperture values within the estimated uncertainties.  
Final photometric errors were computed from the sky noise, 
including corrections for pixel correlations induced by the 
reduction process, and with an additional term representing 
shot noise uncertainty on the net sky background.  

Photometry in the ACS $BViz$ images was measured through 
0\farcs9\ diameter apertures centered at the position of HUDF-JD2 
derived from the NICMOS $H_{160}$--band image, and uncertainties were
estimated based on the background shot noise (which is uncorrelated
in the ACS HUDF data).  The object is undetected (with $S/N < 2$)
in all bands.

The IRAC photometry was measured with SExtractor using 
4\arcsec\ diameter circular apertures, and transformed to total 
magnitudes using aperture corrections based on Monte Carlo 
simulations in which artificial images of compact galaxies 
(half--light radii $\leq 0\farcs5$, appropriate for 
this object) were added to the IRAC images after convolution by
the appropriate PSF and recovered by SExtractor.  The results
were the same within the expected errors when using smaller
(3\arcsec diameter) circular apertures or SExtractor MAG\_AUTO 
measurements.    The three faint galaxies close to HUDF-JD2 lie within 
the aperture used for IRAC photometry. However, these are expected to be too
blue to make any substantial contribution to our estimated IRAC magnitudes.  
The baseline IRAC photometric errors were 
estimated from the same simulations in order to take into
account the statistical effects of crowding in the data.
The errors from the simulations ranged from 0.06 to 0.11 
magnitudes in the four channels.  In practice, we set a floor 
to the IRAC errors of 0.10 magnitudes in channels 1 through 3
and 0.15 magnitudes in channel 4 to also account for remaining 
uncertainties in the IRAC photometric calibration.

 HUDF-JD2 is also detected in the 24$\mu$m MIPS image, as illustrated in 
 Figure~2.  Emission from the foreground spiral galaxy $7\farcs3$ to the 
 southeast makes it necessary to carefully deblend the two objects.
 We have done this in two independent ways, using two independent reductions 
 of the MIPS image.  In one case, we fit point sources to the 24$\mu$m data
 at positions defined by the presence of sources in the higher-resolution
 IRAC images.  We calculate the photometric uncertainties by quadrature
 summation over the residuals after the point source fits are subtracted.
 In this way, we measure 
 a 24$\mu$m flux density of $51.4 \pm 4.0 \mu$Jy.  We also measured photometry
 through small circular apertures centered at the position of the IRAC 
counterpart 
 to HUDF-JD2, applying corrections to total flux computed from an in-flight
 24$\mu$m PSF available from the Spitzer Science Center.  The smallest 
 aperture that we used (4\arcsec\ diameter) minimizes contamination from 
 the neighboring spiral, but requires the largest aperture correction 
 (a factor of 4.8).  This yielded a flux density of $60 \pm 5.0 \mu$Jy, while
 larger apertures were brighter, indicating increasing contributions
 from the neighboring spiral.  We adopt the PSF-fitting 
 flux, but increase the uncertainty estimate to 20\% (10$\mu$Jy)
 to reflect the difficulty of deblending these sources and to encompass
 the range of values measured by the different techniques. 
 The MIPS cataloging procedure will be presented in detail in 
 Chary et al.\ (in preparation).

The photometric data used to construct the spectral energy distribution
discussed in the next section are presented in Table 1.  The difference
between the NICMOS (JH) and IRAC (8$\mu$m) HUDF-JD2 magnitudes in Table 1
with those presented in Yan et al (2004) is mainly due to using differently
reduced NICMOS images, removal of the three nearby faint sources and 
the bright foreground
galaxy in the field of HUDF-JD2 and different photometric procedure 
in the case of the IRAC data. 

\section{Spectral energy distribution and stellar population modeling}

The spectral energy distribution of HUDF-JD2 is presented in Figure 3; 
the ACS measurements are shown as $2\sigma$ upper limits.  
The SED displays a significant change of slope between the measurements 
in the mid--infrared (3.6-8$\mu$m) wavelength range and those in 
the near--infrared (1.1--2.2$\mu$m).  This change of slope is
punctuated by a ``break'' observed between the $K_s$ and 3.6$\mu$m points
mentioned earlier. The SED declines shortward of 1.6$\mu$m and is 
undetected in the HUDF ACS images.    At the longest wavelengths, 
there is another change of slope in the SED between the IRAC and MIPS bands, 
with an upturn at  $\lambda > 8\mu$m (see Figure 6).

 As we will show in the next section, the photometry 
 of HUDF-JD2 at wavelengths shorter than 8$\mu$m can be modeled as 
 emission from an evolved stellar population seen at $z \approx 6.5$. 
 The main evidence for this is the observed $K_s - 3.6\mu$m break which is 
interpreted as the $\lambda_0 3646$\AA\ Balmer break from an
established stellar population.  
 We will also consider old or dust-reddened stellar population models
 at lower redshifts.  These models are illustrated by the spectra
 overplotted in Figure 3.  In any scenario, the measured 24$\mu$m 
 flux density exceeds that predicted by the stellar population models
 alone by more than an order of magnitude.  This, and the distinct
 change of SED slope between the IRAC and MIPS bands, strongly 
 suggest that the 24$\mu$m emission from HUDF-JD2 results from a 
 different physical mechanism than that which is responsible for the 
 bulk of the light at wavelengths shorter than 8$\mu$m.   This may
 arise from warm dust, complex molecules such as polycyclic aromatic 
 hydrocarbons (PAHs), or may be a highly dust-reddened continuum source.
 We will therefore separately model the data at $\lambda \leq 8\ \mu$m
 (\S 4.1 and 4.2) and at 24$\mu$m (\S 4.3).  Any solution must provide 
 an acceptable explanation for the 24$\mu$m flux without invalidating
 the fits obtained for shorter wavelengths.

%%%
\subsection{Stellar Population Modeling}

We now perform detailed modelling of the observed HUDF-JD2 SED in the range
$\lambda \leq 8\ \mu$m, over which the flux is dominated by stellar
component.    
We used two evolutionary synthesis codes in order to model the observed SED:
Starburst99 (SB99: Leitherer et al.\ 1999; V\'{a}zquez \& Leitherer 2004), 
and Bruzual \& Charlot models (BC03: Bruzual \& Charlot 2003).  The SB99 
code has recently been extended to follow stellar populations to ages 
$\sim$5 Gyrs (V\'{a}zquez \& Leitherer 2004). SB99 only supplies 
instantaneous and continuous star formation modes.  For the BC03, 
we used models with exponentially decreasing star formation rates parameterized 
by a time scale $\tau$.   A model with $\tau = 0$ is thus identical to 
the instantaneous star formation in the SB99 models.  
% Both models incorporate contributions from Asymptotic Giant Branch stars, 
% which could contribute to the observed red color.  
For all models we used a Salpeter IMF
with a lower and upper mass cut-off at 0.1 and 100 M$_{\odot}$ respectively.

We searched for the best model fit to the observed data by
simultaneously optimizing the parameters of redshift ($z$), extinction 
($E_{\mathrm{B-V}}$), starburst age ($t_{sb}$) and metallicity ($Z$). 
The continuous star formation mode in the SB99 models could not produce an 
SED even remotely resembling the observed one, and was not considered 
further. In the case of the BC03 models we also include the e-folding time scale for a decreasing 
star formation rate ($\tau$).  The best fit parameters were found through 
$\chi^2$ minimization (with 2 degrees of freedom- 7 data points and 
5 parameters to fit), where we allowed all 
possible combination of model 
parameters in the range; $0 < z < 10$, $0 < E_{\mathrm{B-V}} < 1$, 
$0.1 < t_{sb} < 2.4$ Gyrs and four levels of metallicity $Z$=0.004, 0.008, 0.02 and 0.05.  
For the  BC03 models we used $\tau$=0, 100, 200, 300, 400, 600, 800 and 1000 Myrs.
We used the Madau (1995) prescription for the mean neutral hydrogen 
opacity of the IGM, and modeled dust extinction as a foreground obscuration 
screen at the redshift of the galaxy using the obscuration law of 
Calzetti et al.\  (2000).
%%%This wasn't done!
%With the BC03 models, we also tried the 
%prescription for dust extinction provided by Charlot \& Fall (2000).
As we will see, however, the best--fitting models have little or no 
dust extinction, so the choice of prescription for the reddening makes
little difference on the final results.

The parameters of the best fit models using the two different evolutionary 
synthesis codes are summarized in Table~\ref{table2}.   The results are
very similar for both the SB99 and BC03 models, and we plot the best fit
result from BC03 in Figure 3, together with the observed photometry.

When fitting broad band colors, there are degeneracies between 
parameters such as age, metallicity, extinction, and redshift.
However, with seven significant photometric detections plus four 
upper limits, we find that there are sufficient data to limit 
most of these degeneracies.  In Figure 4, we plot (as a grey scale)
the minimum $\chi^2_\nu$ as a function of redshift and extinction, 
after marginalizing over the other free parameters.   The global 
$\chi^2_\nu$ minimum is marked by an asterisk at $z=6.6$ and 6.5 for the
SB99 and BC03 models, respectively, 
and $E_{\mathrm{B-V}} = 0$ for both sets of models.  The best
fitting redshifts are comparable to those of the most distant
galaxies now known, although we note that the region of acceptable 
$\chi^2_\nu$ values is skewed towards still larger redshifts.  
The parameters not explicitly shown in Figures 4, such as 
starburst age and metallicity, do not vary much over the region 
defined by the 95.4\% confidence level.  Hence, only a small 
number of parameter combinations actually produce good fits
of the model SED, and the redshift turns out to be quite robustly
constrained, as also shown through Monte Carlo simulations (\S4.2). 
The stellar mass of the best--fitting models lies in the range 
$M_\ast = 5-7\times 10^{11} M_{\odot}$.  This represents 
a remarkably large mass for a high redshift galaxy,
and we consider its implications in some detail in \S6.  

The metallicity of HUDF-JD2
is not strongly constrained by the modeling.  This can be seen
by the fact that the BC03 and SB99 models yield different
values for the metallicity, although most other derived
parameters are very similar (Table 2).  Moreover, given the low
metallicity ($Z=0.004$) predicted by the BC2003 model, no gas and a 
high mass ($\sim 5\times 10^{11}$ M$_\odot$), the object will not lie
on any known mass-metallicity relation. To further explore this, 
we fix metallicity
to a relatively high value of $Z=0.02$ (solar) and perform the fit 
to optimise other
parameters. The best-fit gives $z=6.5$, $E_{B-V}=0.0$, $t_{SB}=600$ Myr
and $\tau = 0$ with $\chi^2_\nu = 2.3$. Acceptable fits can also be found for
$t_{SB}=500$ Myr combined with small amount of extinction 
($E_{B-V}\sim 0.03-0.05$). As the age becomes progressively younger 
($<500$ Myr), there is an increased need for higher extinction with
$\chi^2_\nu$ becoming monotonically worse. This is further confirmed by our
Monte Carlo simulation results presented in \S4.2. 

   Figure 4 suggests that a dusty galaxy at $z \approx 2.5$ also
   provides a possible, although less likely, alternative fit to
   the data.  The best-fit parameters corresponding to this model
   are also listed in Table 2, and the model is compared to the
   observed photometry in Figure 3b.  In order not to violate
   the very stringent ACS detection limits, this dusty model must
   also be a post-starburst object, without ongoing star formation.
   A foreground dust screen obeying the Calzetti et al.\ (2000)
   attenuation law cannot adequately suppress the UV flux from active
   star formation while simultaneously matching the near-infrared and
   IRAC photometry, although we note that steeper UV extinction
   laws might be able to accomplish this.  If the 24$\mu$m flux
   from HUDF-JD2 is due to the presence of active star formation
   and not an AGN, then this may appear to contradict the
   ``post-starburst'' nature of the fit to the ACS-through-IRAC
   SED.  However, more complex models with a highly obscured
   starburst or active nucleus plus a more lightly obscured,
   non-star-forming host galaxy, could conceivably fit the data.  
Table 2 also gives parameters for another alternative 
model, which represents an unreddened, passively evolving older 
stellar population at $z = 3.4$.  This gives a still worse fit to 
the data, poorly matching the break from the near-infrared to IRAC 
photometry, and violating the optical non-detection limits in some 
ACS bands.  A similar model dominated by old stellar population was also
used by Yan et al (2004), deriving the same redshift of $z=3.4$ for 
the HUDF-JD2. However, this substantially overpredicts the observed K-band flux 
of the object and seems very unlikely.
  
\subsection{Monte Carlo Simulations}

To test the stability of the parameter values from the model 
fitting, and to better understand the relative likelihoods of the 
high and low redshift interpretations of HUDF-JD2, we have carried 
out Monte Carlo simulations, simultaneously varying the fluxes in all 
bands under the assumption that their errors are normally distributed 
and uncorrelated.  

We generated 1000 realizations of the data and fit each using the BC03 models.
The distribution of the fitted values defines a probability distribution
for each parameter, and is a better assessment of the most likely values. 
In Figure 5 we show the probability distributions for the redshift, stellar 
mass, extinction, and formation redshift ($z_{\mathrm{form}}$) values.  
The $z_{\mathrm{form}}$ is calculated from the estimated photometric redshift
and the age of the stellar population ($t_{sb}$) in HUDF-JD2, using the
cosmology adopted here.    
The redshift distribution shows the same behaviour depicted in Figure 4,
with a median probability around $z\sim 6.9$ but skewed to somewhat higher 
redshift.  As before, a less prominent peak is evident at $z=2$ to 3, 
corresponding to a dust--obscured solution.  
In each panel of Figure 5, the parameter histograms corresponding to 
the low--redshift ($z < 4$) and high--redshift ($z > 4$) solutions are 
shaded differently so that the properties of each family of solutions 
can be easily distinguished.  The distribution for extinction is 
strongly peaked at 
$E_{\mathrm{B-V}}\approx 0.0$, with a small number of high 
values corresponding to the low--redshift solutions.  For the 
high--redshift solutions, the distribution of stellar mass peaks 
around $5 \times 10^{11}$ M$_{\odot}$, as for the best fit model.  
The starburst age (not shown in Figure 5) is evenly spread between 
$\sim 0.4-1.0$ Gyr, where the upper limit corresponds to the best-fit 
solution in the previous section.  The e-folding time is confined 
to 0.0 and 100 Myr, i.e., short compared to the ages of the models, 
while the preferred metallicity is 0.004-0.008.  Considering only
the high-redshift ($z > 5$) solutions, 95\% of the Monte Carlo realizations 
have $z_{\mathrm{form}} > 9$ while 53\%  have $9 < z_{form} < 20$.  
14\% of models exceed the age of the Universe at their best--fitting 
photometric redshifts and hence, are unphysical. 

The relatively narrow distributions for key parameters show that the 
stellar population model fits are stable and are not greatly affected by 
the photometric uncertainties.  In particular, the redshift is one of the 
more robust parameters.  Age and extinction vary only by small amounts, 
and most of the best-fitting models have relatively small amounts of 
ongoing star formation, i.e., with age $t_{sb} \gg \tau$.  Old stellar 
populations at $z < 4.5$ do not fit the data well.  Dusty, post-starburst 
models at $z \approx 2$ to 3 provide relatively poor fits to the observed 
data, outside the 95.4\% confidence range, but do populate a minority of 
the solutions in the Monte Carlo simulations when the photometry is allowed 
to vary within its nominal range of errors.   

In summary therefore, Figures 4 and 5 suggest that the HUDF-JD2 is likely
to be an extremely massive galaxy observed at $6 < z < 8$ 
which formed the bulk of its stars at $z_{\mathrm form} > 9$. 
The size of the observed $K_s - 3.6\mu$m break implies a post-starburst 
system now being observed in a quiescent state. The fraction of realisations 
in Figure 5, leading to low-$z$ ($z < 5$) and high-$z$ ($z>5$) solutions 
over the entire $\chi^2$ range are found to be 15\% and 85\%  respectively, 
as estimated from Monte Carlo simulations, 
with no strong dependence on the explored $\chi^2$ range.

\subsection{Interpreting the 24$\ \mu$m emission}

  We now consider the 24$\mu$m ``upturn'' in the SED, in the 
 context of the stellar population models discussed in the
 previous subsections.  For either of the two classes of
 solutions ($z \sim 6.5$ or $z \sim 2.5$),  the 24$\mu$m 
 emission could be due to the presence of dust heated by star 
 formation or by an active galactic nucleus (AGN).  
 
 For the $z \approx 6.5$ scenario, we regard the AGN hypothesis 
 as being more likely.  At that redshift, the MIPS 24$\mu$m band 
 samples rest-frame wavelengths of $\sim 3.2\mu$m.  Although
 there is a relatively weak PAH emission band at this wavelength,
 the SEDs of star-forming galaxies typically reach a local minimum 
 at roughly this wavelength, between the Rayleigh-Jeans portion 
 of the stellar photospheres and the mid-infared emission from warm 
 dust and PAHs.  Instead, dust-obscured AGN can be much brighter 
 in these wavelengths, due to the presence of a substantially warmer
 dust component heated by the active nucleus.  The reddened
 continuum from the nucleus itself can also contribute to the
 emission at these wavelengths.  Moreover, we have inferred
 that HUDF-JD2 is an exceptionally massive galaxy if it is indeed
 at $z \approx 6.5$.  If the relation between bulge mass and black 
 hole mass, found locally, holds at high redshift, then we might
 expect HUDF-JD2 to host a supermassive black hole at its core.

 Although there is no unique prescription for the near-to-mid-infrared
 spectral energy distribution of AGNs, we illustrate two possible scenarios
 here based on the properties of nearby, infrared-luminous AGNs, namely
 Markarian 231 and NGC 1068. The former is an ultraluminous infrared source, 
optically classified
 as a broad absorption line QSO with an active nucleus bright at
 optical through mid-infrared wavelengths. The latter, instead, is 
 the prototypical Seyfert 2 galaxy, whose active nucleus is highly 
 obscured at optical and near-infrared wavelengths.  We compiled 
 photometry for both objects using the NASA Extragalactic Database.
 Mrk~231 is more distant (178 Mpc) than NGC 1068 (16 Mpc), and most 
 of the available photometry covers larger apertures (10-15 arcsec 
 diameter) encompassing light from both the active nucleus 
 and the host galaxy.  Therefore, there is probably some stellar 
 component to the reported Mrk~231 photometry that we use here;  a hint of this can 
 be seen in a ``bump'' in its SED at rest-frame $\lambda \approx 1.6\mu$m,
 which is a characteristic feature of starlight.  For NGC~1068, we use the 
 smallest aperture nuclear data available, compiled by 
 Galliano et al.\ (2003),  
 and have spliced in an ISOPHOT-S spectrum at 5.8--11.6$\mu$m, 
 whose intensity we have rescaled to match 
 that of the small-aperture photometry in the overlapping wavelength range of 
 (7.7 to 10.4$\mu$m).  This, therefore, is our best
 approximation to the nuclear spectrum of a highly reddened AGN
 in the rest-frame wavelength range 1--10$\mu$m.  We shift both
 AGN templates to $z = 6.5$, and scale them to match the measured 
 24$\mu$m flux of HUDF-JD2.  For our adopted cosmology, 
 this required multiplying Mrk~231 by a factor of 2.85, while
 the NGC~1068 nuclear template was multiplied by a factor of 105.
 Figure 6$a$ shows these two AGN models overplotted with the 
 photometric data and the best-fitting $z = 6.5$ BC03 model from Table 2 and
Figure 3a.  
 
 Both AGN models provide a long wavelength ``upturn'' over the 
 stellar SED for HUDF-JD2.  
 The NGC~1068 nucleus is very heavily reddened, and would contribute 
 negligibly to the flux of HUDF-JD2 at any observed wavelength 
 shorter than 10$\mu$m.  For the rescaled Mrk~231 template, the 
 contribution to the observed IRAC fluxes for JD2 ranges from 
 25\% at 3.6$\mu$m to 53\% at 8.0$\mu$m.  However, we reiterate 
 that the SED adopted here for Mrk~231 most likely includes some 
 component of the stellar host galaxy.  At wavelengths shorter than 
 3$\mu$m, where the flux from HUDF-JD2 drops sharply below the ``break,'' 
 the fractional contribution from the AGN becomes larger again, 
 although we do not have rest-frame UV photometry of Mrk231 suitable for
 extending this comparison across the whole observed wavelength range.  
 
 If an active nucleus contributes some fraction of the light 
 seen at observed 1 to 8$\mu$m wavelengths, this might reduce the implied 
 stellar mass 
 and possibly change other derived stellar population properties.  
 Assuming that the AGN spectral energy distribution is like that of
 the Mrk~231 template, and that it contributes negligibly at shorter
 wavelengths, we subtract the AGN contribution (which is presumed to 
 produce all of the observed 24 $\mu$m emission) from the observed fluxes 
 from the $K_s$-band to 8$\mu$m, and repeat the $\chi^2$ fitting as 
 before.  As a result, we derive a somewhat higher redshift ($z = 7.2$) and a 
 stellar mass that is reduced by a factor of 2.  This is mainly due to
the enhanced strength of the $K_s - m(3.6\ \mu$m) ``break'' 
 although we note that this depends in detail on the behavior of the 
 adopted AGN template in the rest-frame UV.   If, instead, the 
 AGN contribution to the 24$\mu$m emission is represented by the 
 rescaled NGC~1068 nuclear template, it would contribute no significant
 amount of light to fluxes at the observed 1 to 8$\mu$m wavelengths, 
 and correspondingly
 there would be no change to the derived stellar population properties 
 of the host galaxy.

 AGNs commonly emit X-rays, but HUDF-JD2 is undetected in the 1~Msec
 Chandra GOODS-S data.  The nuclear X-ray emission from both Mrk~231 
 and NGC~1068 are believed to be absorbed by a Compton-thick screen
 of neutral gas.  Braito et al.\ (2004) have directly detected the 
 unabsorbed X-ray emission from Mrk~231 at energies up to 60~keV.
 They derive a neutral column density $N$(HI)$ = 2\times 10^{24}$~cm$^{-2}$,
 and infer an intrinsic (unabsorbed) 2-10~keV luminosity of 
 0.5-$2.0\times 10^{44}$~erg~s$^{-2}$.  For a source at $z = 6.5$, 
 the Chandra hard band (2-8~keV) is sensitive to rest-frame energies 
 15-60~keV, where Mrk~231 would be less heavily absorbed.  Extrapolating 
 the luminosity derived by Braito et al.\ with a power law spectrum 
 with photon index $\Gamma = 1.7$ (where $dn/dE \propto E^{-\Gamma}$), 
 and multiplying the luminosity of Mrk~231 by a factor of 2.85 as 
 was done to match the 24$\mu$m photometry, we would predict an X-ray 
 flux of 4.5 to 18$\times 10^{-16}$ erg~s$^{-1}$ in the Chandra
 2-8~keV band. This is 1.9 to 7.5 times larger than the $3\sigma$ 
 hard band detection limits.  However, AGNs may span a broad range of 
 mid-infrared to X-ray flux ratios, depending on their degree of 
 obscuration.  Matt et al.\ (2000) find that the direct X-ray emission 
 from the NGC~1068 nucleus is completely absorbed at all energies out 
 to 100~keV by an absorbing column density $N(HI)\ > 10^{25}$~cm$^{-2}$.
 Only a reflected component of nuclear X-ray emission is detected, 
 about 2 orders of magnitude fainter than the intrinsic nuclear 
 luminosity.  X-rays from a high-redshift AGN of this type would 
 therefore be undetectable with Chandra.

 In the $z \approx 2.5$ hypothesis, the MIPS 24$\mu$m band samples 
 the rest frame at $\sim 7\ \mu$m, where there are strong complexes 
 of PAH emission.  Both starburst and AGN scenarios are plausible
 at this redshift.  However, the absence of detectable X-ray 
 emission from HUDF-JD2 leads us to concentrate on the starburst 
 interpretation here, although again a Compton-thick AGN cannot
 easily be ruled out.  Indeed, Steidel et al.\ (2004) and Donley 
 et al.\ (2005) have shown examples of broad-line or radio-detected
 AGNs in GOODS-N which are faint or undetected in the 2 Msec Chandra 
 X-ray data.   Figure 6$b$ shows the previous best-fit $z \sim 2.5$ 
 solution, a dusty post-starburst spectrum, superimposed 
 with an infrared spectral template from Chary \& Elbaz (2001; CE01), 
 normalized to match the 24$\mu$m measurement.  The model is an 
 ultraluminous infrared galaxy (ULIRG) with 
 $L(8-1000\mu {\mathrm{m}}) = 2 \times 10^{12}$ L$_{\odot}$.
 The CE01 templates include a stellar host galaxy component which 
 dominates the emission at rest-frame wavelengths shorter than 
 about 4$\mu$m, and which we neglect here because we model the starlight
 independently.  The warm dust and PAH emission from the starburst 
 have no significant effect on the photometry in the IRAC bands or
 at shorter wavelengths.

 In summary, we conclude that the 24$\mu$m emission from HUDF-JD2 can
 be interpreted under either the $z \approx 6.5$ or $z \approx 2.5$
 scenarios, and cannot be used as a diagnostic to clearly discriminate
 between these alternatives.  At high redshift, a luminous but highly
 obscured AGN (a ``type II QSO'') is the favored interpretation,
 while at $z \approx 2.5$, either an ultraluminous starburst or an obscured 
 AGN could match the 24 $\mu$m data.  We note that although the BC03 stellar 
 population model, which best fits the ACS through IRAC photometry at 
 $z \approx 2.5$, is heavily reddened, it is not actually forming stars 
 at the time it is observed, but is instead a ``post-starburst'' object
 whose ultraviolet light is fading away.  The 24$\mu$m detection would
 instead imply active star formation with a luminosity $> 10^{12} L_\odot$,
 which would have to be completely hidden by dust at shorter wavelengths.

\section{Spectroscopic Observations}

The main result from the previous section is that HUDF-JD2 is
either an unusually massive galaxy at $z \approx 6.5$, or a dusty
galaxy at $z \approx 2.5$.  In both cases, the best-fitting
models suggest a post-starburst (i.e., evolved) stellar population,
in which case we might expect the galaxy to have very weak or no
emission lines in its spectra.  On the other hand, the detection
at 24$\mu$m would imply the presence of dust-obscured star formation
or an active nucleus.  In order to attempt to measure a redshift
for HUDF-JD2 and to further constrain its possible nature, we have
searched for spectral features (or their absence) in optical and
near-infrared spectra taken with a variety of telescopes
and instruments.  We describe these observations here.

HUDF-JD2 was observed for 4.8 hours with FORS2 on the VLT/UT1 
in service observing on several occasions from UT 2005 February 18
through March 4, using a 1\arcsec\ slit width and the 600z holographic 
grism.  The observations covered the wavelength range 0.749$\mu$m to 
1.075$\mu$m, corresponding to $5.15 < z(\mathrm{Ly}\alpha) < 7.84$, 
with spectral resolution $R = \lambda / \Delta\lambda \approx 1400$.  
No significant emission features or continuum were detected.  Based 
on the measured background noise and simulations adding artificial
emission lines to the data, we estimate that detection flux limits for 
an unresolved emission line falling in between the strong OH lines are 
$2-4 \times 10^{-18}$~erg~s$^{-1}$~cm$^{-2}$ in the range $\lambda\lambda 
0.912- 0.973  \mu$m corresponding to $z(\mathrm{Ly}\alpha) = 6.5-7.0$.

At wavelengths corresponding to Ly$\alpha$ at $z > 6$,
the detection of quite prominent emission lines may be hindered by
the forest of strong OH night sky lines.  Therefore, we have
also analyzed the HST/ACS G800L grism images available 
from the GRAPES project (Pirzkal et al.\ 2004), which provides
continuous wavelength coverage with a smooth sky background
at low spectral resolution of $\sim$80\AA.  The observations were 
carried out at four different position angles to facilitate disentangling 
overlapping spectra in the slitless data, with a total exposure 
time of 25.6 hours.  A spectrum was extracted over a spatial width 
of 0\farcs25 at the nominal position of HUDF-JD2, covering the 
wavelength range 0.55--1.05$\mu$m.  Again, no features were seen 
at $> 3\sigma$ level, with a 3$\sigma$ limiting emission line flux 
limit of $3 \times 10^{-18}$ erg cm$^{-2}$ s$^{-1}$
over the range 9000 to 9600\AA. 

At near-infrared wavelengths, HUDF-JD2 was observed for 2 hours 
with NIRSPEC on Keck II on UT 2005 January 17 in clear skies 
and 0.8--1\arcsec\ FWHM seeing.  The observations used the low
resolution mode with the 75 line/mm grating and a slit width
of 0\farcs76, for a net spectral resolution $R \approx 1500$.
The observations covered the wavelength range 0.947-1.12 $\mu$m 
($6.79 < z(\mathrm{Ly}\alpha) <8.21$), complementing that
examined with FORS2.  Again, no spectral features
were detected.  Over most of the range, the mean 3$\sigma$ 
limiting flux for an emission line with rest--frame velocity 
FWHM~=~300~km~s$^{-1}$ from a point source is 
$5.0\times 10^{-18}$~erg~s$^{-1}$~cm$^{-2}$ (varying from 
$5\times 10^{-18}$ to $1\times 10^{-17}$~erg~s$^{-1}$~cm$^{-2}$ 
depending on the proximity to OH sky lines).  

Finally, the object was also observed at yet longer wavelengths with the 
Gemini South telescope. In a 4.8 hour exposure with GNIRS
we used the short blue camera in cross-dispersed mode, which 
simultaneously covers the $zJHK$ bands with continuous coverage from 
0.94$\mu$m to 2.53$\mu$m.   The 32 line/mm grating and a slit width of 
0\farcs45 gives a net spectral resolution $R \approx 1100$. Again, 
no convincing emission lines were detected.  The observed GNIRS 3$\sigma$
median limits in $zJHK$ bands, representing sensitivity limit of 50\% of 
the spectrum in each order, are: $1.8 \times 10^{-18}$, 
$1.4 \times 10^{-18}$, $3.5 \times 10^{-18}$ and 
$3.1 \times 10^{-18}$~erg~s$^{-1}$~cm$^{-2}$ respectively. 

Collectively, these data provide various upper limits to the ongoing
star formation rate and the contribution from AGN as discussed in \S4.3. 
For the favored high redshift hypothesis, assuming
a Salpeter IMF, the FORS2, NIRSPEC and GNIRS data set limits for the detection
of unabsorbed Lyman~$\alpha$ emission correspond to star 
formation rates of 1-5 $M_{\odot}$ yr$^{-1}$ in the redshift range 
$5.2 < z < 8.2$, consistent with the quiescent state of the stellar 
population implied by the red UV continuum. Also, there are a number of
emission line features originating from AGNs, which lie in the FORS2 and
GNIRS wavelength range covered by our spectroscopic observations (eg. CIV, 
HeII, CIII]). Although weak, non-detection of these lines is consistent with
a highly obscured AGN in HUDF-JD2 at $z\sim 6.5$.

The longer wavelength coverage of the Gemini GNIRS data allows us to
   test the alternative hypothesis that HUDF-JD2 is a dusty galaxy whose
   24$\mu$m emission is produced by an ultraluminous infrared starburst
   or AGN.  At $1.9 < z < 2.8$, H$\alpha$ is shifted into the wavelength
   range covered by the GNIRS $K$-band observations.  At $z = 2.3$, the
   $3\sigma$ GNIRS line flux detection limit corresponds to a luminosity
   limit for redshifted H$\alpha$ of $9\times 10^{40}$~erg~s$^{-1}$.
   This luminosity is quite typical for H$\alpha$ emission from local
   ULIRGs:  $>$55\% of the ULIRGs in  the sample of Armus et al.\ (1988)
   have H$\alpha$ luminosities larger than this value.  
   [OIII] 5007\AA\ lies within the GNIRS $H$-- and $K$--band spectral
   range for $1.9 < z < 4.0$, with a gap around $z \approx 2.8$ due
   to atmospheric absorption.  This line can be as strong as
   the H$\alpha$ in dusty, red galaxies at these large redshifts
   (van Dokkum et al.\ 2005), and may be expected to be particularly
   strong if the 24$\mu$m emission is due to the presence of 
   obscured AGN.  In practice, the spectroscopic limits on emission line
   detection vary substantially with wavelength (and hence, with
   redshift) due to the strong atmospheric OH emission lines and
   water absorption bands.  The non-detection of emission lines in
   the infrared spectra therefore sets a useful, if not definitive,
   limit on the possibility that HUDF-JD2 is a dusty starburst galaxy
   or AGN at lower redshift. Moreover, signatures of Ly$\alpha$ and CIV
   are seen in dusty sub-mm and AGNs at $z\sim 2.5$ (Chapman et al 2003). 
   If HUDF-JD2 resembles these dusty galaxies at their respective redshifts, 
   we would then expect to detect these lines in its spectra. 

\section{Discussion}

 As described in \S4, the photometry of HUDF-JD2 is best fit by 
a two-component model consisting of a passive stellar population 
with an age of several hundred million years and little dust reddening,
observed at $6 < z < 7.5$ (95\% confidence interval), and an AGN producing
the light at longer (24 $\mu$m) wavelengths.      
The redshift is large but not  unprecedented:  other galaxies and quasars 
have been spectroscopically confirmed at similarly large redshifts, 
although we note that the photometric redshift probability distribution 
(Figures 4 and 5) extends up to $z \approx 8.5$.  

The most striking aspect of HUDF-JD2 in this hypothesis, however, is
the large implied stellar mass and the significant age implied by the photometric 
data, particularly from Spitzer/IRAC.  Unlike Ly$\alpha$ emitters or 
UV--bright LBGs which have been discovered at $z > 6$, HUDF-JD2 appears 
{\it not} to be forming stars rapidly, but rather is fading after an earlier 
starburst episode that took place at much higher redshift, $z_{\mathrm{form}} > 10$. 
The implied stellar mass for a Salpeter IMF, approximately 
$6 \times 10^{11} M_\odot$, is four times more massive than 
that of an $L^\ast$ galaxy in the local universe 
($M_\ast = 1.4\times 10^{11} M_\odot$ for a Salpeter IMF and 
our adopted cosmology
; Cole et al.\ 2001).  It is about 50
times greater than that of a typical ($L^\ast$) LBG  
at $z \approx 3$ ($M_\ast \sim 10^{10} M_\odot$; 
Papovich et al.\ 2001) and is even large compared to the 
masses that have been estimated for redder, more evolved 
galaxies from infrared--selected samples at $z \approx 2$ to 3 
($M_\ast \sim 10^{11} M_\odot$; Franx et al.\ 2003;  
Daddi et al.\ 2004;  Yan et al.\ 2004; Labb\'e et al.\ 2005).  
   The implied stellar mass may be reduced somewhat if an obscured
   active nucleus, which is presumed to produce the 24$\mu$m emission,
   contributes significantly to the flux in the IRAC bands.
   The model considered in \S4.3, using a redshifted analog
   to Mrk~231, would reduce the derived stellar mass by about
   a factor of two, whereas the redshifted NGC~1068 model would
   have virtually no impact on the derived host galaxy properties.  

The estimated high mass is also interesting in other respects. 
Although clearly resolved ($\S$3), one might expect HUDF-JD2
to have a larger angular diameter due to a mass-radius relation
(Trujillo et al 2004), extrapolated to $z\sim 6.5$. In fact, we find
this to lie on a reasonable
extrapolation of the radius-redshift relation (Ferguson
et al (2004)), derived from much less massive sample of Lyman Break galaxies. 

Indeed, evidence for established stellar populations at $z\simeq$6-7 is 
becoming stronger from deep Spitzer data. 
Recently, a strongly lensed galaxy was discovered with a likely 
redshift $z \sim 6.8$ (Kneib et al.\ 2004), and with a clear detection 
by Spitzer at 3.6 and 4.5 $\mu$m (Egami et al.\ 2005).  Its SED is 
consistent with a post-starburst galaxy with an age of $\sim 50-200$ Myrs.
Compared to that object, HUDF-JD2 has redder colors and a stronger Balmer 
break, indicating an older age (400--1000 Myrs).  It is also 
approximately 100$\times$ more luminous in the rest-frame $B$--band 
(observed 3.6$\mu$m), after taking into account the lensing 
amplification estimated by Kneib et al.  The redder SED of HUDF-JD2 
also implies a larger stellar mass--to--light ratio ($M/L$), and thus 
a total stellar mass that is $\sim 500$ to $1400 \times$ larger than 
that estimated by Egami et al.\

A number of unlensed $z \approx 5$  to 6 $i_{775}$--dropout LBGs have 
also been detected in the GOODS  IRAC data (Eyles et al.\ 2005; Yan 
et al.\ 2005).  These objects are 
5 to 30$\times$ fainter than HUDF-JD2, with bluer  optical-IR colors 
that imply active (or very recent) star formation and smaller stellar $M/L$.
Their estimated stellar masses, again for a Salpeter IMF, range from 
0.5 to $5\times 10^{10} M_\odot$.  The model fits to HUDF-JD2 suggest 
that it is at least an order of magnitude more massive, and that its 
active star formation phase ended several hundred years prior to 
the redshift at which it is observed.  Finally, we note that the six 
$z_{850}$--band dropout candidates reported by Bouwens et al.\ (2004)
as galaxy candidates at $z = 7$ to 8 are all much fainter than
HUDF-JD2, or are undetected, in the GOODS IRAC 
images.\footnote{Two of the Bouwens et al.\ (2004) objects are close 
enough to brighter foreground galaxies to make the IRAC measurements 
difficult without sophisticated deblending, but would have been easily 
visible if they were as bright as HUDF-JD2.  At least two of the 
Bouwens et al.\ objects, UDF~033238.79-274707.1 and 033242.56-274656.6, 
appear to be individually detected in the GOODS IRAC 3.6$\mu$m image, 
but at very faint fluxes.}

Given the remarkably large stellar mass that we infer for HUDF-JD2
at $z \approx 6.5$, and in the absence of definitive spectroscopic 
confirmation for that redshift, it is prudent to consider alternative 
hypotheses.  These might include an AGN contribution to the light,
gravitational lensing, changes to the stellar population models 
(e.g., to the IMF), or the possibility that the object is at 
$z < 6$, possibly an Extremely Red Object or even an unusual, very red star.

We already explored the effect of an AGN on the estimated mass of
HUDF-JD2 (\S4.3) and concluded that this could reduce its mass by 
$\sim 2$.  Concerning the hypothesis of gravitational lensing,   
HUDF-JD2 is located $7\farcs3$ away from a foreground spiral 
galaxy at $z = 0.457$ (D.\ Stern, private communication).  
At this large impact parameter from an isolated (non-cluster) 
galaxy, we would expect little or no amplification from gravitational 
lensing, nor do we see evidence for elongation or multiple imaging 
that might suggest strong lensing.  It also seems unlikely that 
HUDF-JD2 could be a red star within our own galaxy.  
Dickinson et al.\ (2000) considered stellar explanations for 
another ``$J$--band dropout'' object from the Hubble Deep Field North 
in some detail, and found that only heavily dust--enshrouded objects 
like some carbon stars or Mira variables could match the very red 
infrared colors.  Such objects would be far brighter than HUDF-JD2 
unless they were located far outside our own Galaxy, at distances 
of many Mpc.  Moreover, HUDF-JD2 is clearly extended in the 
NICMOS $H_{160}$ images, making a stellar origin unlikely.  

From the stellar population model fitting discussed in \S4
(see Figures 3 through 5), the most likely alternative to the passive, 
$z \approx 6.5$ scenario is that HUDF-JD2 is a heavily dust--obscured,
post--starburst galaxy at lower redshift.  Such models generally
do not match the measured spectral energy distribution as well
through the ``break'' region between the near--infrared and
IRAC wavelengths, resulting in poorer $\chi^2$ overall.  
Additionally, these models require both an absence of 
recent star formation {\it and} heavy dusty extinction
in order not to violate the extraordinarily deep optical
detection limits provided by the ACS HUDF data.
As discussed in \S5, the non--detection of line emission 
in the Gemini GNIRS data can be used to argue against 
a dusty starburst at lower redshift, although at certain
redshifts, line emission might be obscured by atmospheric 
features or fall outside the spectral range of the existing
data. 

The stellar mass inferred from SED modeling could be reduced 
by adopting an IMF that is deficient (relative to Salpeter) in lower--mass 
stars, such as that proposed by Chabrier et al.\ (2003).  However, this 
would not substantially change the mass {\it relative} to that of other 
galaxies at similar or lower redshifts, unless we appeal to an IMF which
varies substantially from galaxy to galaxy or as a function
of redshift.   

Recently, Maraston (2005) has presented
models which include a greater contribution to the red optical 
and near--infrared light from thermally pulsating AGB stars at 
stellar population ages of a few hundred Myr compared to traditional
models such as BC03 or SB99.  The Maraston models have redder  
UV to near-IR colors and smaller $M/L$ at the wavelengths spanned 
by IRAC photometry at $z \approx 6.5$.  For a simple stellar 
population with age 1 Gyr (i.e., equivalent to our best--fitting BC03
or SB99 models), the Maraston models have approximately 30\% 
smaller $M/L$ at rest--frame 1$\mu$m (i.e., roughly the 
wavelength sampled by IRAC channel 4 at $z \approx 6.5$).  
We verified that the observed SED can be reproduced with somewhat
younger ages, reduced formation redshift and stellar mass.
It is unlikely, however, that the mass could be reduced by
more than a factor of 2.

If the $z > 6$ interpretation of HUDF-JD2 is correct, then this 
galaxy formed the bulk of its stars at very high redshift, 
$z > 9$ within a period of a few hundred Myr before entering 
a quiescent phase.  Remarkably, this galaxy would have formed
a stellar mass several times greater than that of our Milky Way 
(even allowing for IMF variations), and would have done so
very rapidly, when the universe was only a few hundred million
years old.  Assuming the formation starburst lasted $< 100$~Myr, 
as implied by the stellar population modeling, the star formation 
rate must have been $> 5000 M_\odot$/year, implying a remarkably
luminous birth event, particularly at observed--frame 
mid--infrared wavelengths if the star formation was relatively
unobscured.  Such a massive galaxy is expected to host a supermassive 
black hole (SMBH) at its center, producing the observed 24 $\mu$m emission 
as discussed in \S4.3. Therefore, at $z > 6$, while HUDF-JD2 is forming
a significant number of stars, the SMBH is also expected to accrete 
a lot of mass.

It is instructive to estimate the mass of dark matter halo required 
to host the stellar mass of HUDF-JD2. Klypin, Zhao \& Somerville (2002) 
estimate the total (virial) and baryonic ($M_{stellar}$ + $M_{gas}$) mass of 
the Milky way as $10^{12}$ M$_\odot$ and $6\times 10^{10}$ M$_\odot$ 
respectively. Assuming the same ratio ($M_{Baryon}/M_{total} = 0.06$)
for HUDF-JD2 and 100\%  efficiency in converting baryons to stars, 
we estimate dark matter halo mass of $ 10^{13}$ M$_\odot$, required
to host $6\times 10^{11}$ M$_\odot$ mass of HUDF-JD2. However, considering
the universal $M_{Baryon}/M_{total}$ ratio of 0.15, the dark matter halo
mass reduces to $4\times 10^{12}$ M$_\odot$. Assuming the Sheth-Tormen
modified Press-Schechter formalism (Sheth \& Tormen 1999) and a $\Lambda$CDM
model then predicts a space density of $\sim 10^{-6}$ Mpc$^{-3}$ for such
dark matter halos (Somerville 2004; Mo \& White 2002) which is significantly 
smaller than that observed here (based on HUDF-JD2 alone). Should such a 
population exist,
they are expected to be strongly clustered. 

Analyzing optical--through--IRAC photometry for red, infrared--selected
galaxies at $z \approx 2$ to 3, Labb\'e et al.\ (2005) and
Yan et al.\ (2004) have argued that some are dominated by at least
$10^{11} M_\odot$ of old stars which must have formed at $z \gg 5$.  
Similar arguments have been found in the literature concerning
``red, dead'' galaxies at almost any redshift, including Extremely Red
Objects at $z \approx 1$ to 2 (see McCarthy 2004 for a review) and
giant elliptical galaxies in the local universe (Eggen, Lynden-Bell \& 
Sandage 1961). 
We estimate the expected colors of HUDF-JD2 when passively evolving
the $z=6.5$ SED fit solution to $z\sim 2.5$, the average redshift of 
the "Distant, Red Galaxies" (DRGs)- (van Dokkum et al 2003; 
Toft et al. 2005) and the upper redshift limit of the BzK selected 
objects (Daddi et al 2004). We predict 
$J-K=2.9$; $i-K=5.6$; $K- m(3.6 \mu m)=0.7$ and $K_s\sim 20$ for HUDF-JD2 when 
evolved to $z=2.5$. This agrees closely with the observed colors of the
DRGs, estimated as $\langle J-K_s \rangle\ > 2.3$; 
$\langle i-K_s \rangle\ = 5$ and $\langle K_s - m(4.5\ \mu m)\rangle = 0.9$ 
(Labbe et al 2005). Moreover, given the predicted $K_s$-band magnitude, 
such evolved objects will be detected in the K20 or BzK surveys.     
HUDF-JD2 might be seen as a progenitor to such galaxies, and a sign 
that at least a few objects may have formed quite large masses of 
stars ``monolithically'' at very early times, and evolved quiescently
down to lower redshifts.  Even if the source did not collapse
monolithically but arose from several 
sub-galactic components which subsequently coalesced, it seems such
early phases should be visible with future large--aperture 
telescopes such as the Thirty Meter Telescope and the James Webb 
Space Telescope.  The mere existence of objects with such large 
star formation rates at very high redshift would have important 
implications for the re-ionization of the intergalactic medium, 
and we discuss this further in a companion paper (Panagia et al.\ 
in preparation).

\acknowledgements{This paper is
based on observations taken with the NASA/ESA {\it Hubble Space Telescope}, 
which is operated by AURA, Inc. under NASA contract NAS5-26555, 
{\it Spitzer Space Observatory} and the W. M. Keck Observatories. 
   Support for this work, part of the Spitzer Space
   Telescope Legacy Science Program, was provided by NASA
   through contract number 1224666 issued by the
   Jet Propulsion Laboratory, California Institute of
   Technology, under NASA contract 1407.  
We acknowledge award of Director's Discretionary Time at the Gemini-South 
observatory and the Very Large Telescope (VLT) at Cerro Paranal, Chile. 
We are grateful to Matt Mountain and Phil Puxley for making the spectroscopic
observations possible. We acknowledge an anonymous referee for carefully
reading the manuscript and for very constructive comments.  
We are gratful to Dave Alexander for his help
in measuring Chandra detection limit and to GRAPES team for the use
of their data.}

\clearpage

\begin{figure}
\epsscale{0.8}
\plotone{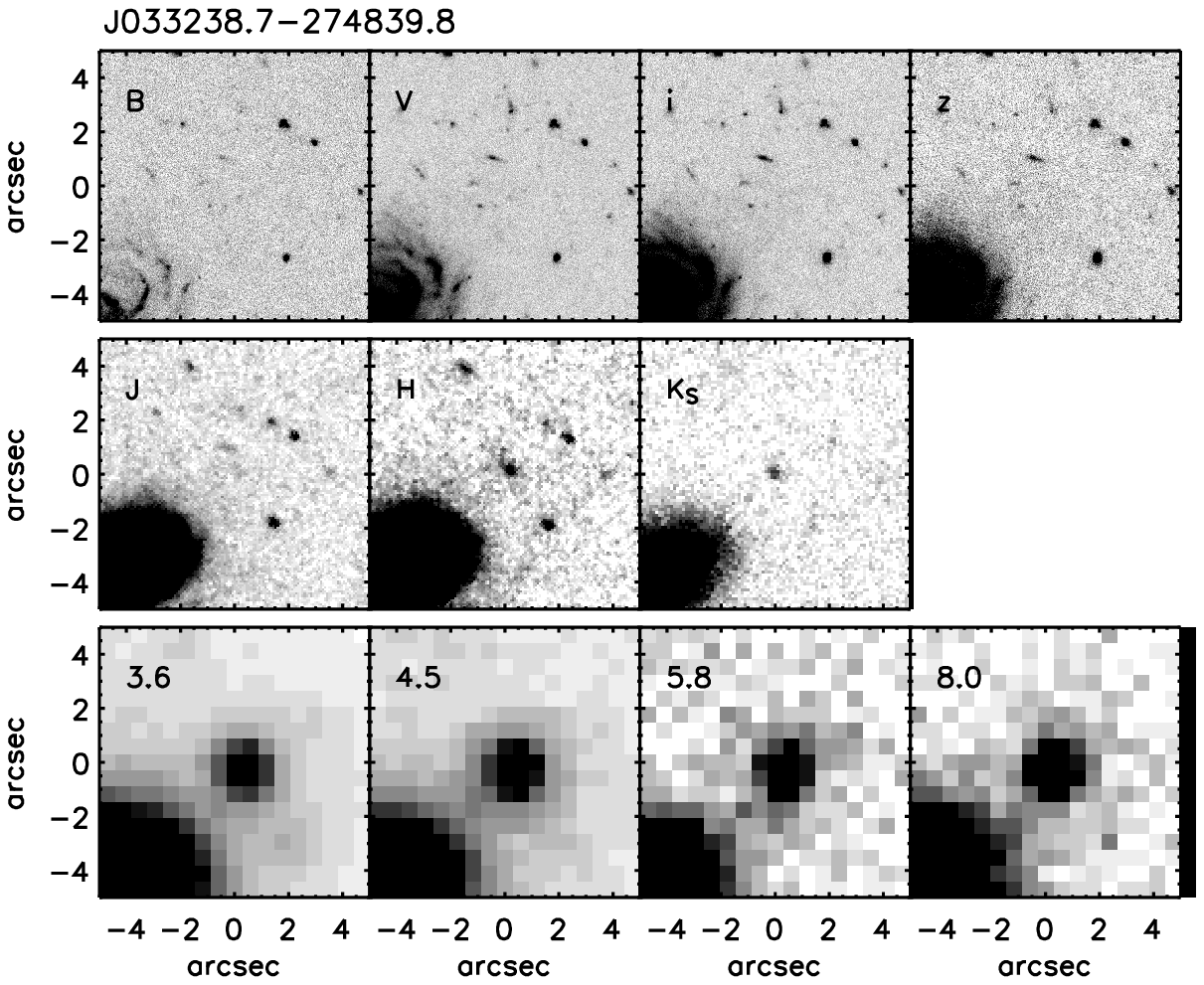}
\caption{Images of the $J$--dropout candidate HUDF--JD2 
($\alpha=$3:32:38.74; $\delta=$ $-$27:48:39.9 J2000) from
HST/ACS ($B_{435}V_{606}i_{775}z_{850}$), HST/NICMOS ($J_{110}H_{160}$), 
VLT/ISAAC ($K_s$) and Spitzer/IRAC (3.6-8.0 $\mu$m). The $K_s$ ISAAC image
is from deep FIRES observations.}

\label{fig1}
\end{figure}

\begin{figure}
%\epsscale{1.0}
\includegraphics[angle=0,scale=0.8]{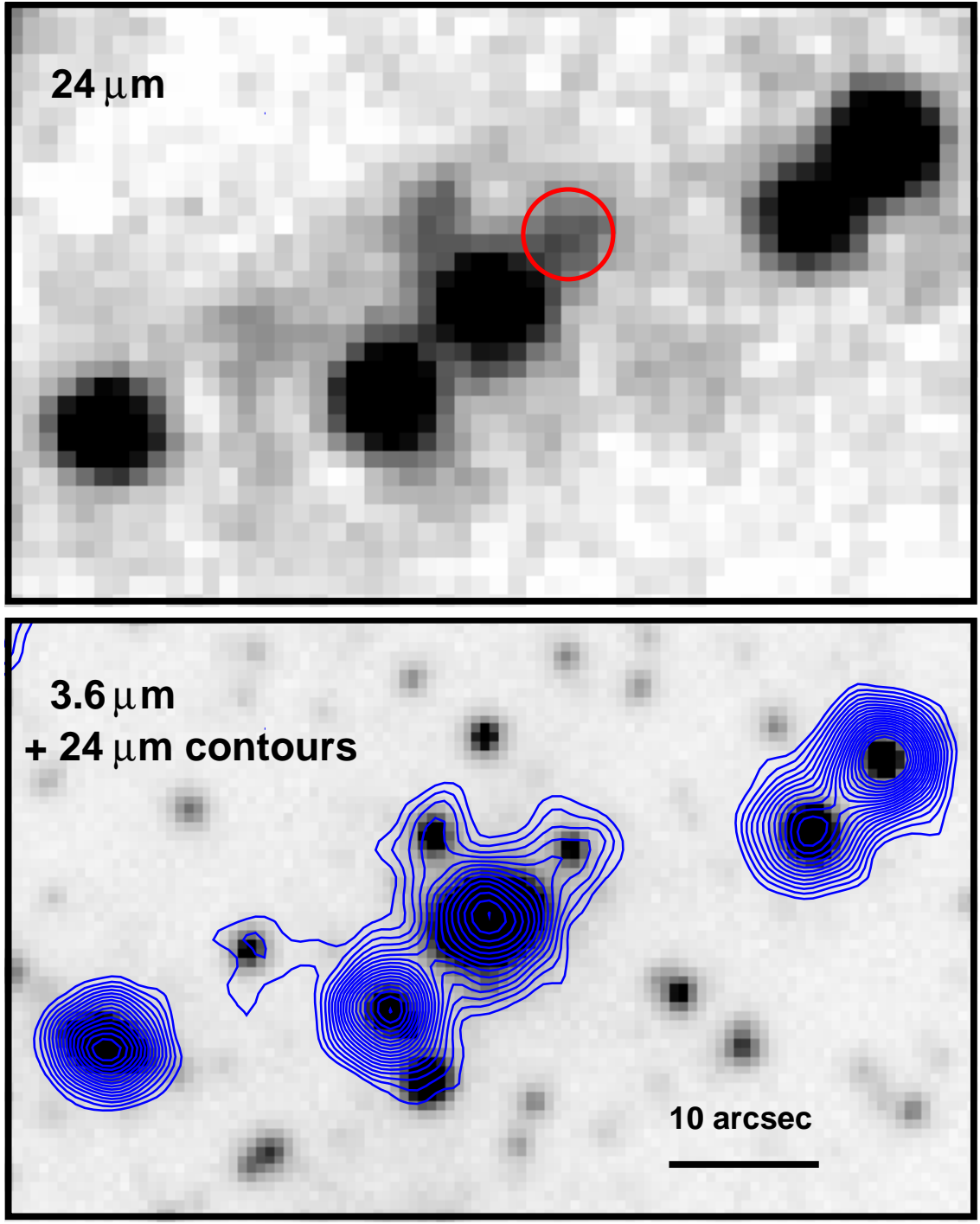}
\caption{{\it Top panel:}  MIPS 24$\mu$m image of the region around HUDF-JD2.
The position of the source as measured from the NICMOS $H_{160}$
image is marked by the red circle, which has a diameter of 6\arcsec,
approximately the FWHM of the 24$\mu$m point spread
function.
{\it Bottom panel:} IRAC 3.6$\mu$m image, with contours from the
24$\mu$m superimposed.}

\label{fig2}
\end{figure}

\begin{figure}
%\epsscale{1.0}
%\plotone{sed_dual_new.eps}
\includegraphics[angle=0,scale=0.8]{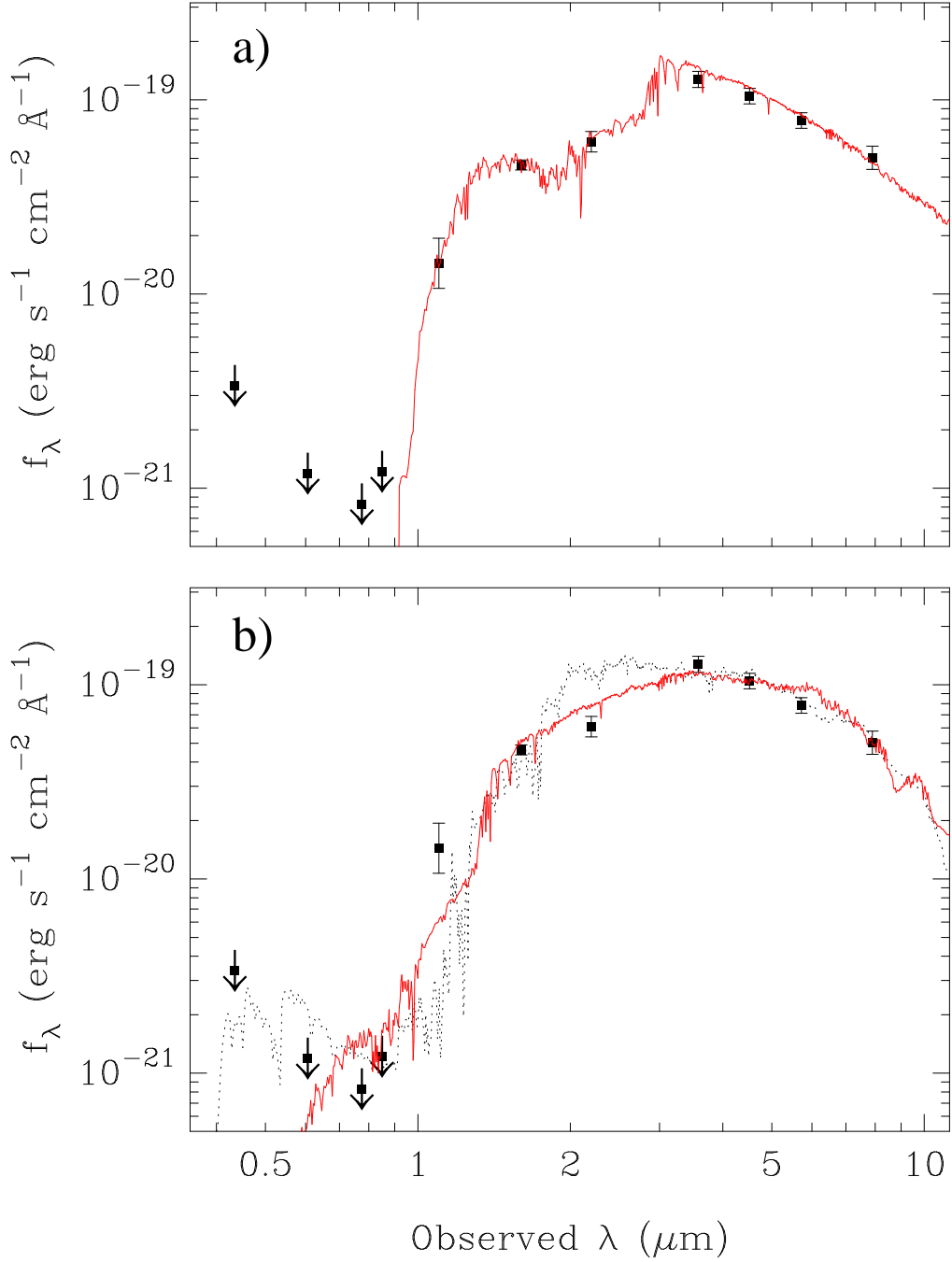}
\caption{Observed and model SED of HUDF-JD2.  {\it Top panel:} The red line is the model 
SED based on the best fit Bruzual \& Charlot model (see Table~\ref{table2}) at
a redshift $z = 6.5$.  {\it bottom panel:} Alternative, lower redshift models.
The solid line shows the best fit for a dusty model galaxy
   at $z = 2.5$, the best fit from the secondary $\chi^2$
   minimum seen in Figure 4.  The parameter values are given
in Table 2.  The dotted line shows the SED for an old
   population with $z = 3.4$ and an age of 2.4~Gyr (Table 2).
(Table~\ref{table2})}
\label{fig3}
\end{figure}

\begin{figure}
\epsscale{1.0}
\plotone{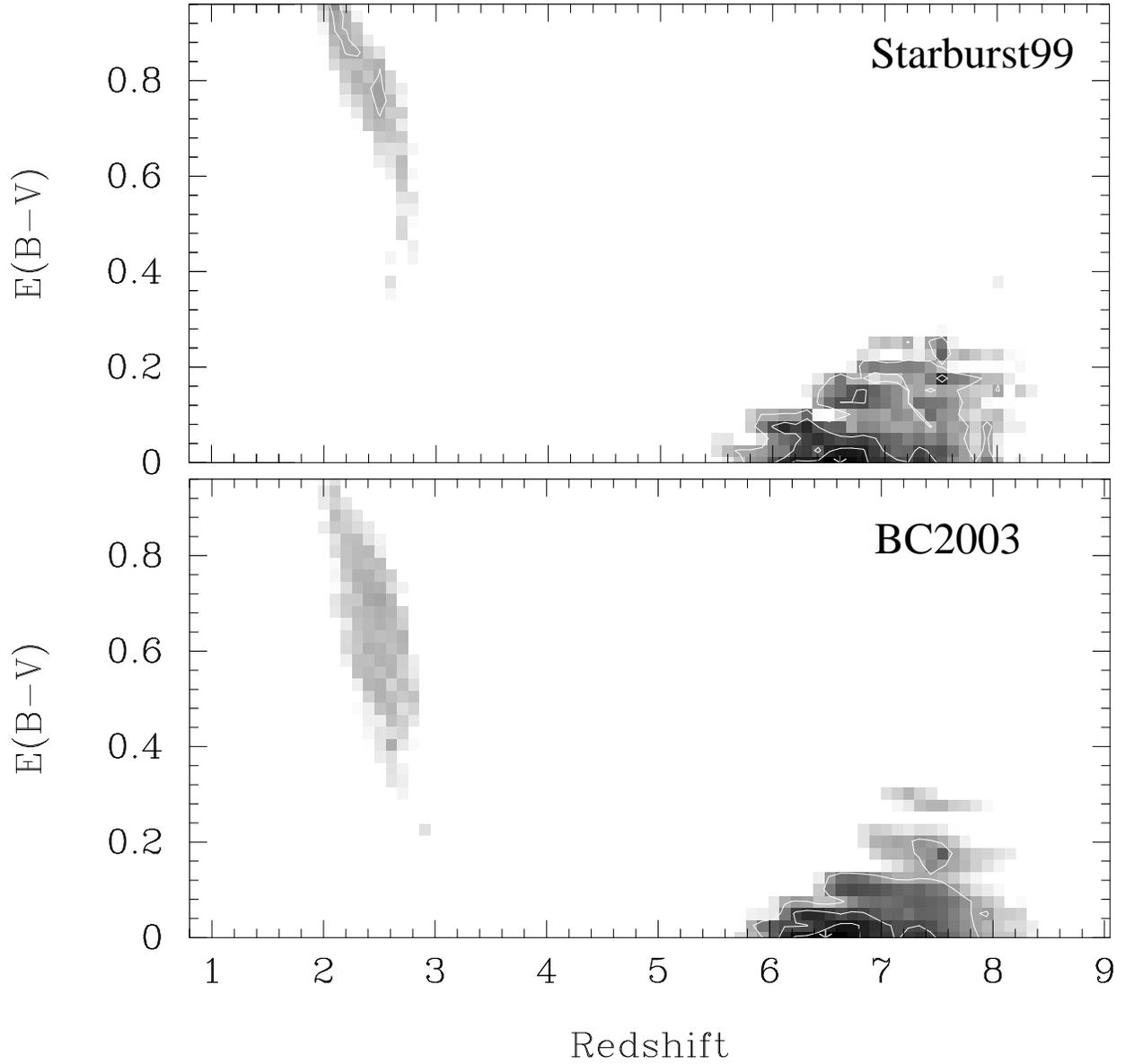}
\caption{The reduced $\chi^2$ values of the best fitting models as a function of 
redshift $z$ and extinction E$_{\mathrm{B-V}}$, shown as grey scale, for 
{\bf top} Starburst99 and {\bf bottom} Bruzual \& Charlot models.  Black 
corresponds to the lowest $\chi^2$ value, $\sim$1.9, and white to $\chi^2 >10$.
The contours show the 68\%, 90\% and 95.4\% confidence of the $\chi^2$ fit, 
and the white asterisks mark the best-fitting solutions.}
\label{fig4}
\end{figure}

\begin{figure}
\epsscale{1.0}
\includegraphics[angle=-90,scale=0.8]{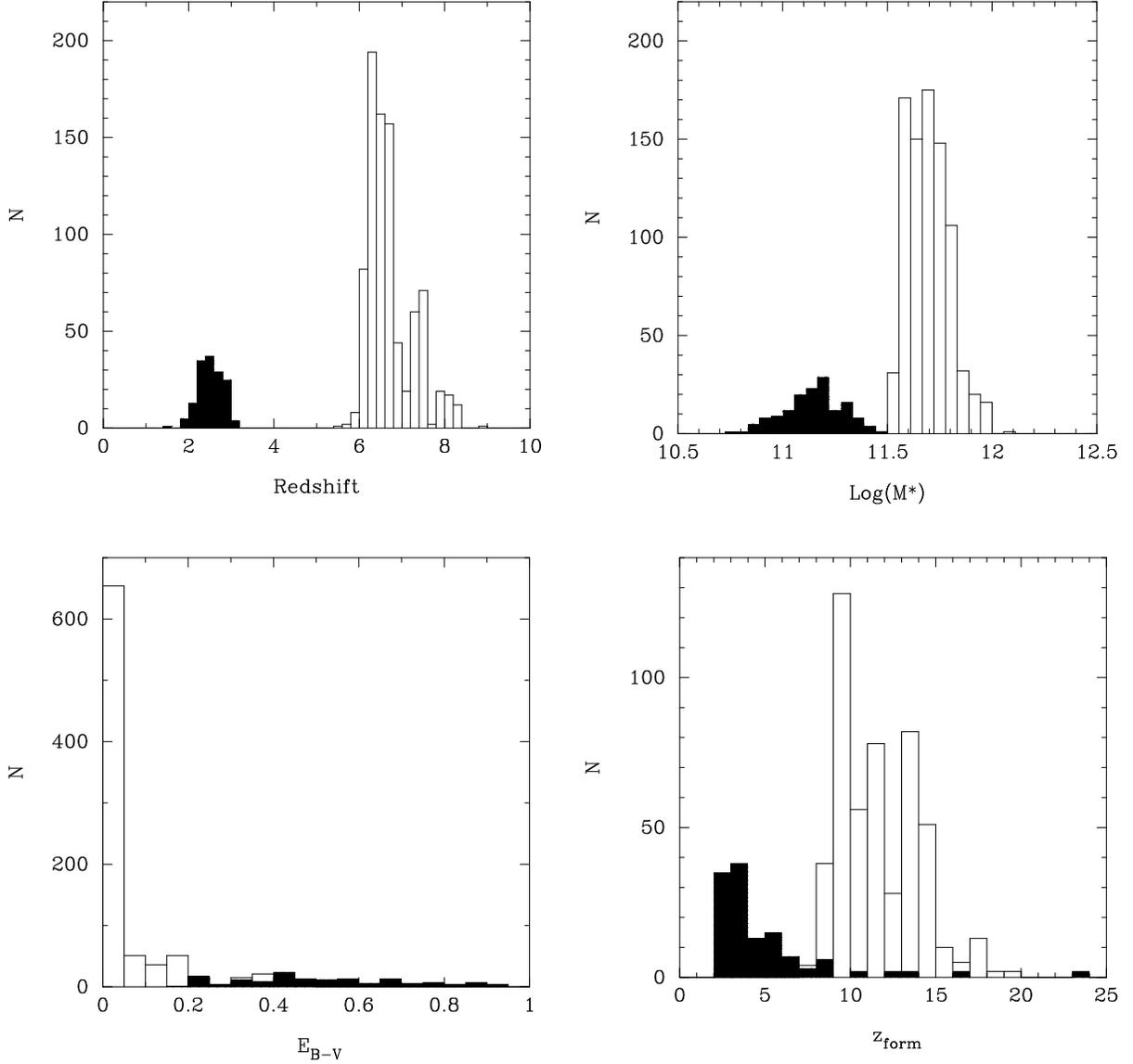}
%\plotone{fig5.ps}
\caption{Results from Monte Carlo simulations where the photometric measurements
of HUDF-JD2 in all bands were allowed to vary according to their estimated errors.
The panels show the distribution of parameter values for the best-fitting 
BC03 models fit to each Monte Carlo realization.  
{\em Upper left:} redshift.
{\em Upper right:} stellar mass.
{\em Lower left:} extinction $E(B-V)$.
{\em Lower right:} Formation redshift $z_{\mathrm{form}}$, where realizations
where the formation redshift is in conflict with the age
of the universe have been removed. Dark histograms correspond to
the alternative dusty solution at $z\sim 2.5$.} 

\label{fig5}
\end{figure}

\begin{figure}
\epsscale{0.8}
\plotone{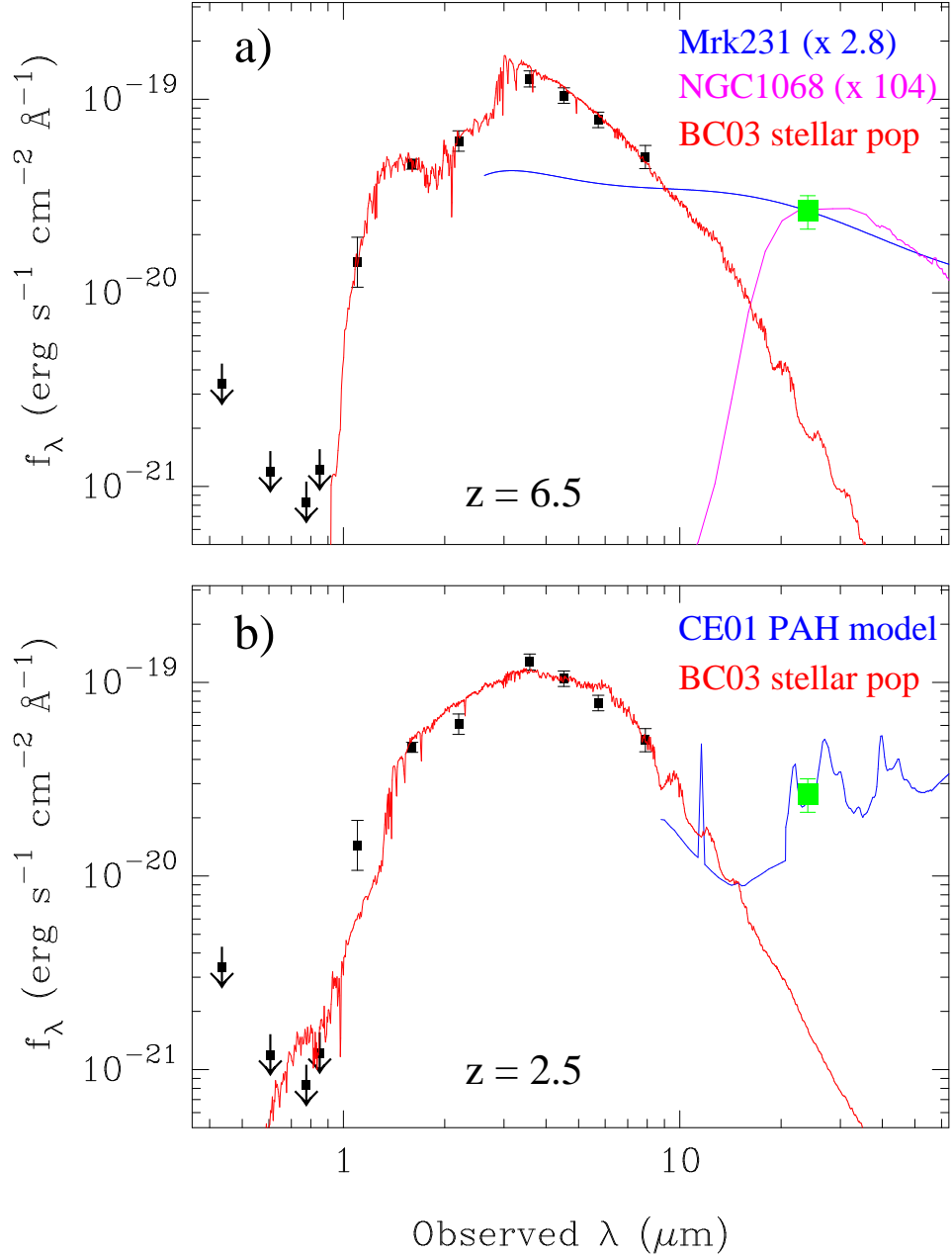}
\caption{Observed and model spectral energy distributions for HUDF-JD2,
   extended to include the 24$\mu$m measurement.  The models
   which best fit the data at $\lambda \leq 8\mu$m (see Figure 3
   and Table 2) are shown as red lines in both panels.
   {\it Top panel:} The $z = 6.5$ solution.  Two models of
   obscured, infrared-luminous AGN are shown, normalized to the
   observed 24$\mu$m point:   the ultraluminous infrared AGN
   Mrk 231, scaled in luminosity by a factor of 2.85, and
   the highly obscured Seyfert 2 nucleus NGC 1068, scaled
   by a factor of 105.
   {\it Bottom panel:} The $z = 2.5$ solution.  A model 
   starburst galaxy template from Chary \& Elbaz 2001, with
   $L_{IR} = 2\times 10^{12} L_\odot$, is shown, normlaized to
   the 24$\mu$m photometry.}
\label{fig6}
\end{figure}

% New version of table, provided by Tommy Wiklind on 2 May.

\clearpage
\begin{table}

\caption[]{Photometry for J- dropout candidate HUDF-JD2 
(RA:53.161425 Dec:--27.81107 (J2000)). Magnitudes, followed in separate line by their associated errors, 
are in the AB system. HST/ACS magnitudes are given as 2$\sigma$ upper limits.}
\begin{tabular}{llllllllllll}
& & & & & & & & & & &\\
\hline
\hline
 $B_{435}$ & $V_{606}$ & $i_{775}$ & $z_{850}$& $J_{110}$ & $H_{160}$&$K_s$& $m_{3.6}$& $m_{4.5}$ & $m_{5.8}$ & $m_{8.0}$ & $m_{24}$ \\
\hline
& & & & & & & & & & &\\
$> 30.61$& $>31.02$ & $> 30.88$ & $> 30.26$ & 27.02 & 24.94 & 23.95 & 22.09& 21.80 & 21.60 & 21.38 & 19.63\\
- & - & - & - & 0.32 & 0.07& 0.13& 0.10 & 0.10& 0.10& 0.15 & 0.21\\
%& & & & & & & & & & &\\
\hline
\end{tabular} \label{table1}
\end{table}
\clearpage
\begin{table}
%\begin{center}
\caption[]{Parameters for the best-fit models to HUDF-JD2}
\begin{tabular}{ccccc}
  &  &  &  &  \\
\hline
\hline
\multicolumn{1}{c}{}                     &
\multicolumn{1}{c}{SB99$^{a)}$}           &
\multicolumn{1}{c}{BC03$^{b)}$}          &
\multicolumn{1}{c}{Dusty$^{c)}$ z$<$4}    &
\multicolumn{1}{c}{Old$^{d)}$ z$<$4}      \\
\hline
z  & 6.6 & 6.5 & 2.50 & 3.4 \\
E$_{\mathrm{B-V}}$  & 0.0   & 0.0 & 0.70 & 0.0 \\
$Z$ & 0.020 & 0.004 & 0.020 & 0.050 \\
t$_{\mathrm{SB}}$ (Gyr) & 1.0 & 1.0 & 0.6 & 2.4 \\
$\tau$ (Myr) & --  & 0 & 0 & 0 \\
\hline
L$_{\mathrm{bol}}$ (L$_{\odot}$) & $1.0 \times 10^{12}$ & $1.0 \times 10^{12}$ &
 $3.5 \times 10^{11}$ & $2.1 \times 10^{11}$\\
M$_*$/L$_{\mathrm{bol}}$ (M$_{\odot}$/L$_{\odot}$)$^{e}$ & 0.71 & 0.42  & 0.29 
& 0.95 \\
M$_*$ (M$_{\odot}$)$^{f}$  & $7 \times 10^{11}$  & $5 \times 10^{11}$ & $1 \times 10^{11}$ & $2 \times 10^{11}$ \\
%\hline
z$_{\mathrm{form}}$$^{g)}$ & $>$9 & $>$9 & $--$ & $--$ \\
%\hline
$\chi^2_\nu$$^h$  & 1.8 & 1.9 & 6.7 & 29.9\\
\hline
  &  &  &  &  \\  
\end{tabular} \label{table2}
%\end{center}
\ \\
a) Starburst99 version 5.0 (V\'{a}squez \& Leitherer 2004) \\
b) Bruzual \& Charlot (2003) \\
c) Best-fit parameters forcing z$<$4 \\
d) Best-fit parameters forcing z$<$4 and $E_{\mathrm B-V} = 0$ \\
e) Stellar mass-loss is included in the ratio\\
f) Salpeter IMF with lower and upper mass cut-off at 0.1 and 100 M$_{\odot}$\\
g) 95\% lower confidence limit based on high-redshift ($z > 4$) solution from
Monte Carlo simulations.\\
h). $\chi^2$ per degree of freedom ($\nu=2$)
\end{table}


\begin{thebibliography}{}

% \bibitem[]{824} Adelberger, K. L. \& Steidel C. C. 2000, ApJ. 544, 218 
\bibitem[]{825} Ajiki, M. et al. 2004, PASJ. 56, 597
\bibitem[]{826} Armus, L., Heckman, T.M., Miley, G.K., 1989, ApJ, 347, 727
\bibitem[]{827} Alexander, D.\ M., et al., 2003, AJ, 126, 539.
\bibitem[]{828} Bertin, E. \& Arnout, S. 1996, A\&AS 117, 393
% \bibitem[]{829} Blanton, M. R., et al., 2001, ApJ, 121, 2358
\bibitem[]{830} Bouwens, R. J.  et al., 2004 ApJ 606, 25
\bibitem[]{8301} Braito, V. et al. 2004 A \& A  420, 79
\bibitem[]{831} Bruzual, G. \& Charlot, S. 2003, MNRAS 344, 1000
\bibitem[]{832} Bunker, A. J. Stanway, E. R., Ellis, R. S. \& McMahon, R. 2004,
MNRAS 355, 374
\bibitem[]{834} Calzetti, D. et al. 2000 Ap. J. 533, 682
\bibitem[]{836} Chabrier, G., 2003, PASP, 115, 763
\bibitem[]{8361} Chapman, S. C.; Blain, A. W.; Ivison, R. J.; Smail, Ian R. 2003, Nature 422, 695
\bibitem[]{835} Charlot, S., \& Fall, S.\ M., 2000, ApJ, 539, 718
\bibitem[]{8361} Chary, R.; Elbaz, D. 2001 Ap.J. 556, 562
\bibitem[]{837} Chen, H.-W., \& Marzke, R., 2004, ApJ, 615, 603
\bibitem[]{838} Cole, S., et al., 2001, MNRAS, 326, 255
\bibitem[]{839} Daddi, E. et al., 2004, ApJ, 617, 746
\bibitem[]{8390} Daddi, E. et al. 2005 ApJL (in press)
\bibitem[]{840} Dickinson, M., et al., 2000, ApJ, 531, 624
\bibitem[]{841} Dickinson, M., Giavalisco, M., and the GOODS Team, 2003, 
in ``The Mass of Galaxies at Low and High Redshift,'' 
eds.\ R.\ Bender \& A.\ Renzini, Springer Verlag
\bibitem[]{844} Egami, E. et al. 2005 ApJ 618, L5
\bibitem[]{845} Eyles, L.\ P., et al., 2005, MNRAS, in press
\bibitem[]{846} Franx, M. et al. 2003 ApJ 587, 79
\bibitem[]{8461} Galliano, E.; Alloin, D.; Granato, G. L.; Villar-Martin, M. 2003, A\&A 412, 615
\bibitem[]{847} Giacconi, R., et al., 2002, ApJS, 139, 369
\bibitem[]{848} Giavalisco, M.  2002 ARA\& A 40, 579
\bibitem[]{849} Giavalisco, M. et al., 2004a, ApJL, 600, 103
\bibitem[]{850} Giavalisco, M. et al., 2004b, ApJL, 600, 93

\bibitem[]{8500} Klypin, A.; Zhao, H. S.; Somerville, R. S. 2002, Ap.J. 573, 597
\bibitem[]{851} Kneib, J., Ellis, R. S., Santos, M. R. \& Richard, J. 2004 ApJ
607, 697
\bibitem[]{853} Labb\'e, I., et al., 2005, ApJ, in press astro-ph 0504219
\bibitem[]{854} Leitherer, C. et al. 1999 ApJS 123, 3
\bibitem[]{855} Loeb, A. \& Barkana, R. 2002, ARA\& A 39, 19
\bibitem[]{856} Madau, P., 1995, ApJ, 441, 18
\bibitem[]{857} Maraston, C., 2005, MNRAS, in press
\bibitem[]{8571} Matt, G.; Fabian, A. C.; Guainazzi, M.; Iwasawa, K.; Bassani, L.; Malaguti, G. 2000 MNRAS 318, 173
\bibitem[]{858} McCarthy, P.\ M.\ J., 2004, ARAA, 42, 477
% \bibitem[]{859} Mo, H. J. \& White S. D. M. 2002 MNRAS 336, 112
\bibitem[]{860} Ouchi, M. et al., 2004 ApJ 611, 660
\bibitem[]{861} Papovich, C. 2001 ApJ. 559, 620
\bibitem[]{862} Pirzkal, N., et al., 2004, ApJS, 154, 501
\bibitem[]{8621} Reddy, Naveen A.; Steidel, Charles C. 2004 ApJL 603, 13
\bibitem[]{863} Rhoads et al., 2004, ApJ, 611, 59
\bibitem[]{8631} Rigby, J. R.; Rieke, G. H.; Perez-Gonzalez, P. G.; Donley, J. L.; Alonso-Herrero, A.; Huang, J.-S.; Barmby, P.; Fazio, G. G. 2005, Ap.J 627, 134
\bibitem[]{864} Somerville et al., 2001 MNRAS 320, 504
\bibitem[]{8640} Somerville R. S. astro-ph 0401570
\bibitem[]{865} Stiavelli, M., Fall, S. M. \& Panagia, N. 2004 ApJ 600, 508
\bibitem[]{866} Steidel, C. C. Adelberger, K. L. Shapley, A. E. Pettini, M. 
Dickinson, M. Giavalisco, M. 2003 ApJ 592, 728
\bibitem[]{868} Thompson, R., et al., 2005, AJ, in press
\bibitem[]{8680} Toft, S. van Dokkum, P., Franx, M. Thompson, R. I., Illingworth, G. D., Bouwens, R. J. \& Kriek, M. astro-ph 050345
\bibitem[]{8681} Trujillo et al. 2004, Ap.J. 604, 521 
\bibitem[]{869} van Dokkum, P. et al. 2004 ApJ. 611, 703
\bibitem[]{8691} van Dokkum, P. et al. 2005 ApJL 622, L13
\bibitem[]{870} V\'{a}zquez, G. A. \& Leitherer, C. 2004, astro-ph/0412491
% \bibitem[]{871} Wiklind, T. \& Henkel, C. 1992 ApJ 123 123
\bibitem[]{872} Yan, H. \& Windhorst, R. A. 2004, ApJL 612, L93
\bibitem[]{873} Yan, H. et al., 2004 ApJ 616, 63
\bibitem[]{874} Yan, H. et al., 2005, ApJ, submitted
\end{thebibliography}
\end{document}